# Selective Direct Bonding of High Thermal Conductivity 3C-SiC Film to *β*-Ga$_2$O$_3$ for Top-Side Heat Extraction

Jianbo Liang[1,2,*], Hiromu Nagai[2], Zhe Cheng[3,*], Keisuke Kawamura[4], Yasuo Shimizu[5], Yutaka Ohno[5], Yoshiki Sakaida[4], Hiroki Uratani[4], Hideto Yoshida[6], Yasuyoshi Nagai[5], and Naoteru Shigekawa[1,2]

[1]Department of Physics and Electronics, Osaka Metropolitan University, 3-3-138 Sumiyoshi, Osaka 558-8585, Japan

[2]Graduate School of Engineering, Osaka City University, 3-3-138 Sumiyoshi, Osaka 558-8585, Japan

[3]Department of Materials Science and Engineering and Materials Research Laboratory, University of Illinois at Urbana-Champaign, Urbana, IL 61801, United States.

[4]The SiC Division, Air Water Inc., 2290-1 Takibe, Toyoshina Azumino, Nagano 399-8204, Japan.

[5]Institute for Materials Research (IMR), Tohoku University, 2145-2 Narita, Oarai, Ibaraki 311-1313, Japan

[6]The Institute of Scientific and Industrial Research (ISIR), Osaka University, 8-1 Mihogaoka, Osaka 567-0047, Japan

[*]Corresponding authors: liang@omu.ac.jp; zcheng18@illinois.edu.

*β*-Ga$_2$O$_3$ is a wide bandgap semiconductor with electrical properties better than SiC and GaN which makes it promising for applications of next-generation power devices. However, the thermal conductivity of *β*-Ga$_2$O$_3$ is more than one order of magnitude lower than that of SiC and GaN, resulting in serious thermal management problems that limit device performance and reliability. This work reports selectively transferring of high thermal conductivity 3C-SiC thin film grown on Si to *β*-Ga$_2$O$_3$ (001) substrate using surface activated bonding (SAB) technique at room temperature, to attempt extracting the heat from the surface of the devices. A 4.5-nm-thick interfacial crystal defect layer is formed at the as-bonded 3C-SiC/*β*-Ga$_2$O$_3$ interface. The thickness of the interfacial crystal defect layer decreases with increasing annealing temperature, which decreases to 1.5 nm after annealing at 1000 °C. No voids and unbonded area are observed at the interfaces, even after annealing at temperature as high as 1000 °C. The thermal boundary conductance (TBC) of the 1000 °C-annealed 3C-SiC/$\beta$-Ga$_2$O$_3$ interface and thermal conductivity of the $\beta$-Ga$_2$O$_3$ substrate was measured by time-domain thermoreflectance (TDTR). The 3C-SiC/$\beta$-Ga$_2$O$_3$ TBC value was determined to be 244 MW/m$^2$·K, which is the highest value ever reported for SiC/Ga$_2$O$_3$ interfaces, due to the high-quality heterointerface. Our works demonstrate that selective transferring of 3C-SiC film to the $\beta$-Ga$_2$O$_3$ substrate is an efficient path to improve heat dissipation of the $\beta$-Ga$_2$O$_3$ power devices.

## ■ INTRODUCTION

Beta-phase gallium oxide ($\beta$-Ga$_2$O$_3$) is a type of semiconductor material with a bandgap (~ 4.9 eV) larger than SiC and GaN which makes it possesses higher electrical breakdown field strength.[1,2] Its Baliga figure-of-merit (BFOM) is several times higher than that of SiC and GaN. Furthermore, large-diameter, potentially low-cost, and high-quality $\beta$-Ga$_2$O$_3$ substrate can be obtained by the floating zone (FZ) and edge-defined film-fed growth (EFG) methods.[3,4] Accordingly, $\beta$-Ga$_2$O$_3$ is a potential candidate for next-generation high-power devices and is being extensively studied for high-power and low-loss device applications.[5-7] However, $\beta$-Ga$_2$O$_3$ has an anisotropic thermal conductivity ranging from 10 to 27 W/m·K, which is at least one order of magnitude lower than other wide bandgap semiconductors such as SiC (320 W/m·K) and GaN (200 W/m·K). When operating at high power, low thermal conductivity will cause serious thermal management problems which degrade the performance and reliability of the device.[8,9] Therefore, an efficient thermal management strategy is the key for real-world $\beta$-Ga$_2$O$_3$ device applications.

To solve the thermal management problems, an effective method is to combine $\beta$-Ga$_2$O$_3$ with high thermal conductivity substrates. $\beta$-Ga$_2$O$_3$-on-diamond structures have been developed by direct growth $\beta$-Ga$_2$O$_3$ on diamond or bonding $\beta$-Ga$_2$O$_3$ to a diamond substrate at low temperature.[10,11] The $\beta$-Ga$_2$O$_3$ grown on the diamond is polycrystalline $\beta$-Ga$_2$O$_3$, which is difficult for scalable manufacturing. For $\beta$-Ga$_2$O$_3$ bonded with diamond, the bonded area is very small and there are lots of cracks formed in the $\beta$-Ga$_2$O$_3$ film bonded to diamond. Recently, the ion-cutting $\beta$-Ga$_2$O$_3$ thin film bonded to 4H-SiC substrate via an Al$_2$O$_3$ interlayer using surface activated bonding (SAB) method has been reported.[12] The thickness of the $\beta$-Ga$_2$O$_3$ film exfoliated by the ion-cutting technique is about 200-400 nm, which is just for $\beta$-Ga$_2$O$_3$ lateral devices. For $\beta$-Ga$_2$O$_3$ vertical devices, a drift layer thickness larger than 10 μm is required.[13,14] Recently, a 21-μm-thick $\beta$-Ga$_2$O$_3$ film-on-polycrystal SiC structure has been developed by bonding the $\beta$-Ga$_2$O$_3$ substrate to a polycrystal SiC substrate, and then thinning the $\beta$-Ga$_2$O$_3$ substrate using mechanical lapping and chemical mechanical polishing methods.[15] However, the thick $\beta$-Ga$_2$O$_3$ film leads to a large thermal resistance itself, which makes bottom-side cooling through substrates difficult due to the low thermal conductivity of $\beta$-Ga$_2$O$_3$. Furthermore, the film with uniform thickness and large area is very difficult to obtain by thinning the substrate process. Therefore, top-side heat extraction and spreading is an effective way to lower the temperature in $\beta$-Ga$_2$O$_3$ devices.

In this work, we selectively bond a high thermal conductivity 3C-SiC film to a $\beta$-Ga$_2$O$_3$ substrate using the SAB technique at room temperature to explore heat extraction from the top side of the $\beta$-Ga$_2$O$_3$ vertical devices. The thermal stability of the 3C-SiC/$\beta$-Ga$_2$O$_3$ interface is tested at 400, 700, and 1000 °C in N$_2$ gas ambient pressure, which is necessary for $\beta$-Ga$_2$O$_3$ device fabrication processes. The annealing process on the effects of the interfacial structure and atomic behavior is systematically investigated by transmission electron microscopy (TEM) and energy-dispersive X-ray spectroscopy (EDS). The thermal boundary conductance of the 3C-SiC/$\beta$-Ga$_2$O$_3$ interface and the thermal conductivity of the $\beta$-Ga$_2$O$_3$ substrate is measured by time-domain thermoreflectance (TDTR).

# RESULTS AND DISCUSSION

A high thermal conductivity 3C-SiC film grown on Si substrate with hole patterns was transferred to a $β$-Ga$_2$O$_3$ substrate in this work, as shown in the schematic diagram of Figure 1. The 3C-SiC film with a thickness of about 1 μm was grown on the Si substrate by low-temperature chemical vapor deposition. The hole patterns were fabricated by photolithography and inductive coupled plasma reactive ion etching (ICP-RIE) processes. The thin film with hole patterns was directly bonded to the $β$-Ga$_2$O$_3$ substrate using the SAB method at room temperature. After bonding, the Si substrate used for 3C-SiC films growth was removed by mechanical polishing and reactive ion etching (RIE) processes. More detailed experimental information can be found in the Experimental Section.

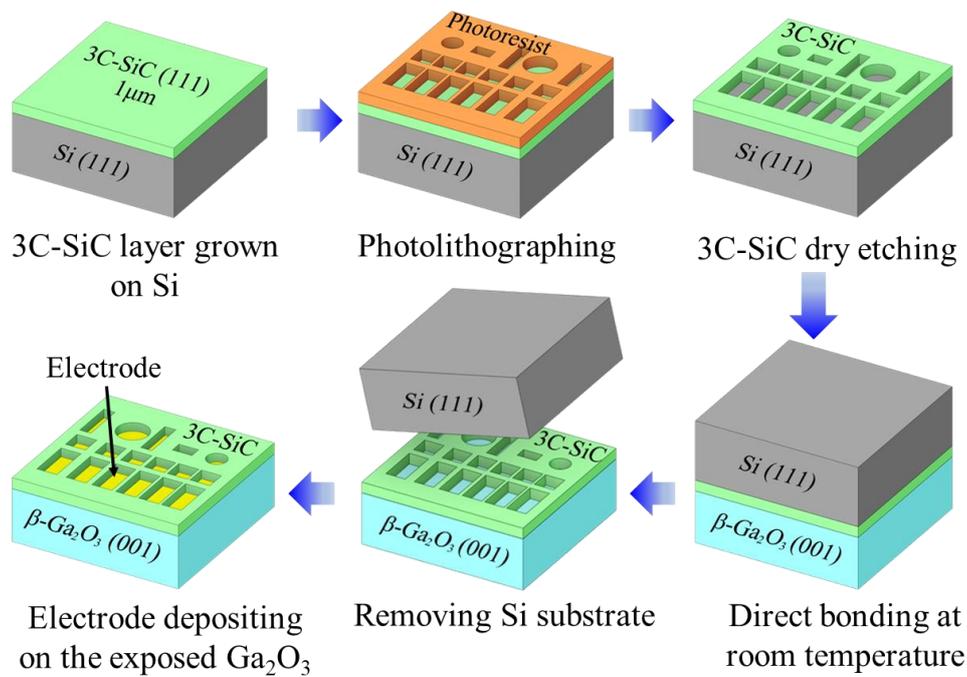

**Figure 1**. Schematic diagram of 3C-SiC thin film selectively bonding to $β$-Ga$_2$O$_3$ substrate. A 3C-SiC thin film with hole patterns transferred to the $β$-Ga$_2$O$_3$ substrate. Electrodes can be deposited on the exposed Ga$_2$O$_3$ surfaces.

An optical microscopy image and a SEM image of 3C-SiC thin film with hole patterns transferred to the $β$-Ga$_2$O$_3$ substrate and a cross-sectional SEM image of an as-bonded 3C-SiC/$β$-Ga$_2$O$_3$ interface are shown in Figures 2(a) - (c), respectively. No unbonded area was observed in the 3C-SiC thin film bonded to the *β*-Ga$_2$O$_3$ substrate, which demonstrated the excellent bonding and transferring of 3C-SiC thin film with hole patterns to the *β*-Ga$_2$O$_3$ substrate. No micro voids were observed at the as-bonded interface. These results show that the high-quality direct bonding of the 3C-SiC thin film and the *β*-Ga$_2$O$_3$ substrate was achieved at room temperature. A thin 3C-SiC film with hole patterns was bonded to the *β*-Ga$_2$O$_3$ substrate, which allows for later electrodes depositing on the exposed *β*-Ga$_2$O$_3$ surface.

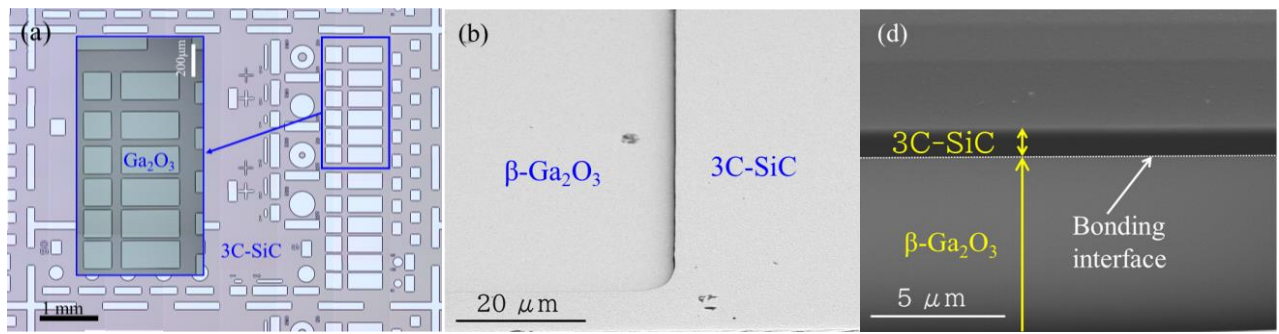

**Figure 2**. (a) Optical microscopy image and (b) SEM image of 3C-SiC thin film with hole patterns transferred to the *β*-Ga$_2$O$_3$ substrate. (c) Cross-sectional SEM image of an as-bonded 3C-SiC/*β*-Ga$_2$O$_3$ interface.

Figures 3(a) and (b), (c) and (d), (e) and (f), and (g) and (h) show high-resolution TEM (HRTEM) images taken along Ga$_2$O$_3$ [100] and SiC [$\bar{1}$01] zone axes of the as-bonded, 400°C-annealed, 700 °C-annealed, 1000 °C-annealed 3C-SiC/β-Ga$_2$O$_3$ interfaces, respectively. There is a thin intermediate layer observed. The thickness of the intermediate layer formed at the as-bonded interface was measured to be approximately 4.5 nm, which decreased from 4.5 to 3.5 nm, from 3.5 to 2.5 nm, and from 2.5 to 1.5 nm after annealing 400, 700, and 1000°C, respectively. The reduction of the intermediate layer thickness highly depended on the post-annealing temperature. It should be noted that some discontinuous lattice fringes were observed in the intermediate layer of the as-bonded interface as shown in Figure3(b). The distribution range of the discontinuous lattice fringes was found to be in the intermediate layer with a thickness of approximately 3 nm, which is originated from the extension of the Ga$_2$O$_3$ lattice fringe. Fast Fourier Transform (FFT) of the intermediate layer reveals that the intermediate layer is composed of two parts: one part composed of a crystal defect layer of Ga$_2$O$_3$, and the other part an amorphous layer. The crystal defect layer thickness decreased with increasing post-annealing temperature. No change was observed in the amorphous layer thickness with increasing annealing temperature. However, after annealing at 1000 °C, the amorphous layer changed into the crystal defect layer composed of 3C-SiC and $β$-Ga$_2$O$_3$. More important, no unbonded area and mechanical defects were not observed near the interface, even after annealing at 400, 700, and 1000 °C. Thus, these results demonstrate that the SiC/Ga$_2$O$_3$ interface has excellent thermal stability and can satisfy the annealing process demand for $β$-Ga$_2$O$_3$ device fabrication.

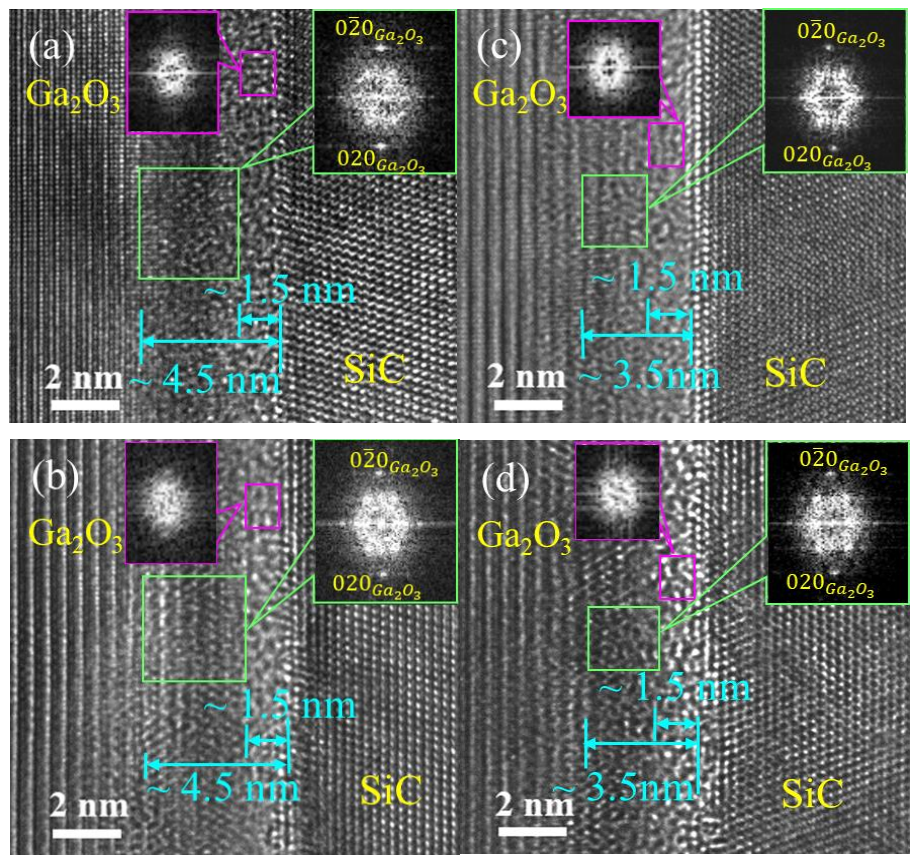

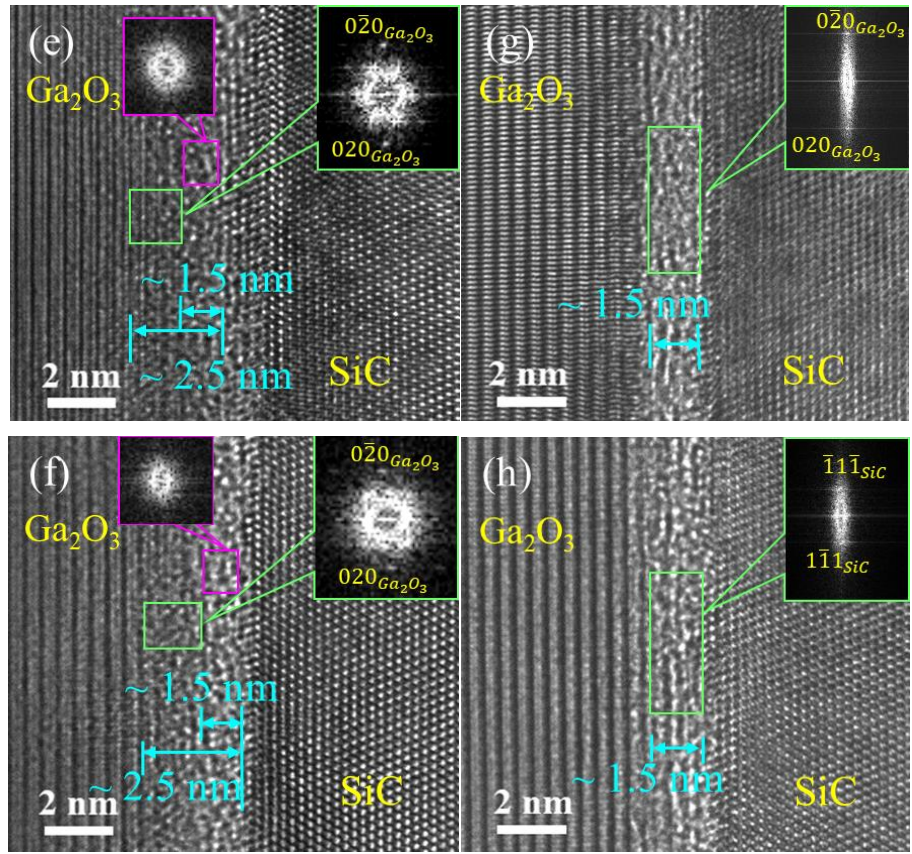

**Figure 3.** HRTEM images taken along Ga$_2$O$_3$ [100] zone axis and SiC [$\bar{1}$01] zone axis on the as-bonded (a) and (b), 400 °C-annealed (c) and (d), 700 °C-annealed (e) and (f), and 1000 °C-annealed (g) and (h) 3C-SiC/β-Ga$_2$O$_3$ interfaces.

The X-ray intensity profiles for Ga, O, Si, C, Ar, and Fe atoms across the as-bonded, 400 °C-annealed, 700 °C-annealed, and 1000 °C-annealed 3C-SiC/β-Ga$_2$O$_3$ interfaces are shown in Figures 4(a), (b), (c), and (d), respectively. The inset STEM images indicate the corresponding location of measured X-ray intensity profiles. Intensity gradients for the Ga, O, Si, and C atoms were observed in the intermediate layer and became more abrupt with increasing annealing temperature. Furthermore, we find that Ga and O, and Si and C atoms diffused into the SiC and Ga$_2$O$_3$ substrates adjacent to the intermediate layer, respectively. However, the diffusion depths decreased with increasing annealing temperature. In addition, a small peak for the intensity profile of the Ar atoms was observed, the intensity reached the background level after annealing temperature higher than 400 °C. A small peak for the Fe atoms was also observed at the as-bonded interface, which should be originated from the metal vacuum chamber during the Ar beam irradiation of the bonding process. Previous research on the compositional analysis of the as-bonded GaAs/Si interface has shown that Fe atoms existed on the surface irradiated by Ar beam irradiation, which indicated that the peak position of Fe atoms' relative intensity was located on the bonding interface.[16] When annealing temperature is higher than 400 °C, no peak for Fe atoms intensity profile was observed, which was attributed to the concentration of Fe atoms distributed in the intermediate layer is lower the resolution of the EDS detector sensor due to Fe atom diffusion after annealing.

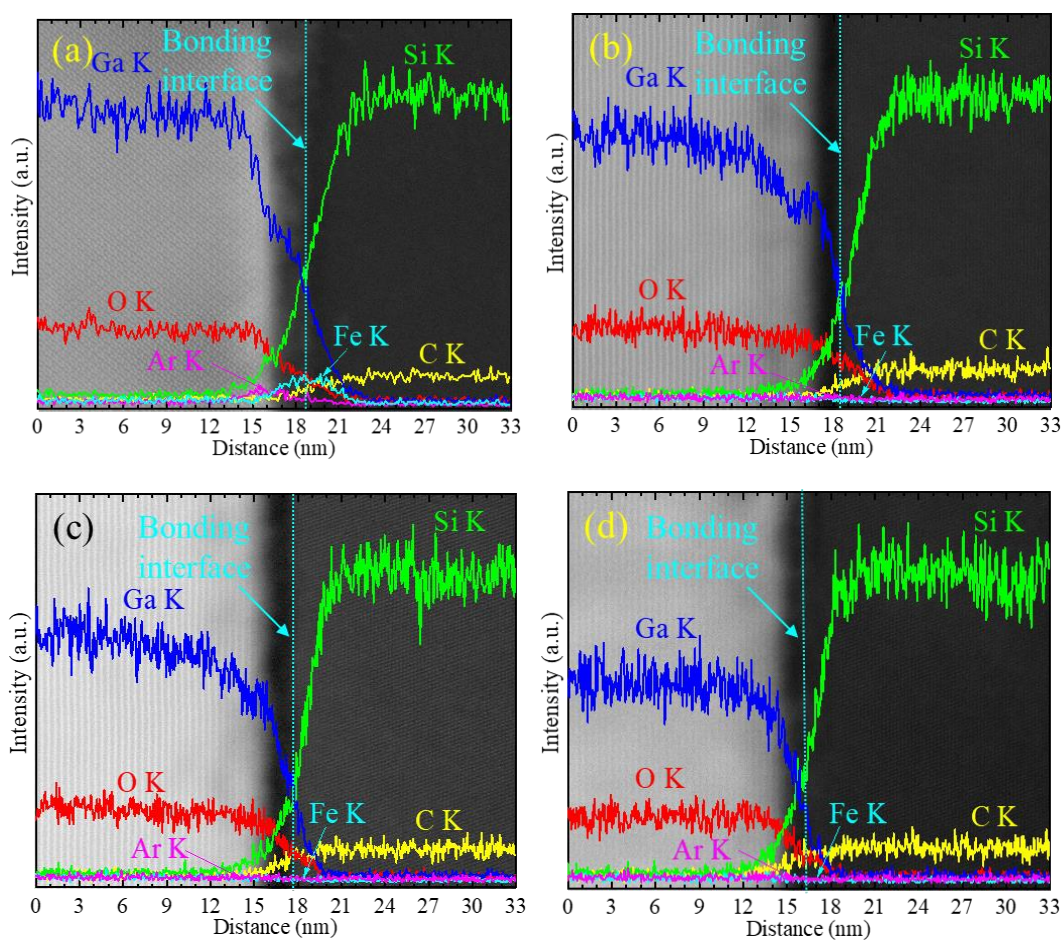

**Figure 4.** X-ray intensity profiles for Ga, O, Si, C, Ar, and Fe atoms (blue, red, green, yellow, purple, and cyan, respectively) across (a) the as-bonded, (b) 400 °C-annealed, (c) 700 °C-annealed, and (d) 1000 °C-annealed 3C-SiC/β-Ga$_2$O$_3$ interfaces. The inset STEM images indicate the corresponding location of the measured X-ray intensity profiles.

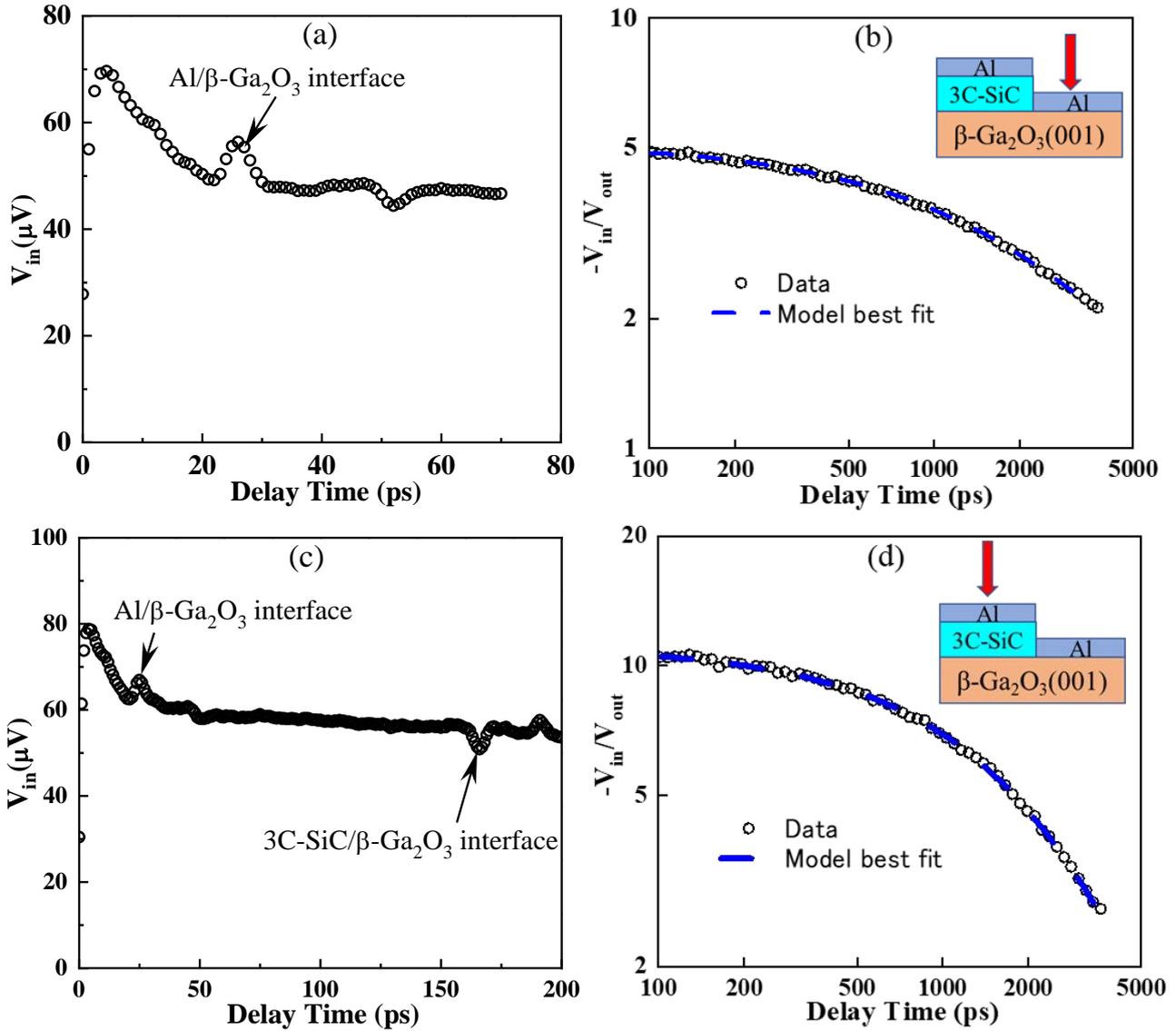

**Figure 5.** TDTR measurements on the 1000 °C-annealed 3C-SiC/β-Ga$_2$O$_3$ sample. (a) Picosecond acoustic measurements on the β-Ga$_2$O$_3$ (001) substrate. (b) The data fitting of TDTR experimental data on the bare β-Ga$_2$O$_3$ (001) substrate. (c) Picosecond acoustic measurements on the 1000 °C-annealed 3C-SiC/β-Ga$_2$O$_3$ interface. (d) The data fitting of TDTR experimental data on the 1000 °C-annealed 3C-SiC/β-Ga$_2$O$_3$ interface.

The thermal conductivity of the β-Ga$_2$O$_3$ (001) substrate without 3C-SiC is measured by time-domain thermoreflectance (TDTR). Figure 5(a) shows the picosecond acoustic echoes of the Al/β-Ga$_2$O$_3$ interface while Figure 5(b) shows the TDTR data fitting of the experimental ratio signal with the analytical heat transfer solution of the sample structure. The measured thermal conductivity of the

β-Ga$_2$O$_3$ (001) substrate is 15 W/m-K, which is used as a known parameter in the data fitting of measurements on the 1000 °C-annealed 3C-SiC/β-Ga$_2$O$_3$ interface. The echo from the 1000 °C-annealed 3C-SiC/β-Ga$_2$O$_3$ interface is observed (Figure 5(c)). The local thickness of the 3C-SiC is determined as 877 nm. Here, the sound velocity of 3C-SiC along the [111] direction was used 12.5 km/s.[17] The symmetric echo shows the high quality of the interface. By fitting the TDTR ratio signal (Figure 5(d)), the thermal boundary conductance (TBC) of the 1000 °C-annealed 3C-SiC/β-Ga$_2$O$_3$ interface is determined as 244 MW/m$^2$·K, which is 53% higher than the TBC of directly bonded 4H-SiC/β-Ga$_2$O$_3$ interface and more than twice as high as the TBC of bonded 4H-SiC/β-Ga$_2$O$_3$ interfaces with Al$_2$O$_3$ interfacial layers.[18,12]

The crystal defect layer formed at the as-bonded 3C-SiC/β-Ga$_2$O$_3$ interface is unlike the Si/4H-SiC,[19] β-Ga$_2$O$_3$/poly-SiC[15], Si/β-Ga$_2$O$_3$,[20] and 4H-SiC/β-Ga$_2$O$_3$[21] interfaces fabricated by SAB technique that there was an amorphous layer formed at the interface. The main reason is that the damage caused by Ar beam irradiation was reduced by lowering the acceleration voltage. The amorphous layer formed at the interface is due to the excess Ar beam irradiation energy. For the crystal defect layer thickness decreased with increasing annealing temperature, a reasonable explanation is that the crystal defect layer was self-healed by the annealing process. Based on reports that the amorphous layer formed at the interface was recrystallized by annealing at high temperatures.[19,22,23] Furthermore, the amorphous layer changed into the crystal defect layer after annealing at 1000 °C is also due to the recrystallization of the amorphous layer. Although the thermal expansion coefficient of β-Ga$_2$O$_3$ (6.34 × 10$^{-6}$ K$^{-1}$) is 81% higher than that of 3C-SiC (3.5 × 10$^{-6}$ K$^{-1}$),[24,25] no interfacial debonding was observed at the interface even after annealing at 1000 °C. This should be contributed to the thin bonded 3C-SiC film and the crystal defect layer formed at the bonding interface that can relax thermal stress caused by the difference in thermal expansion coefficient mismatch between β-Ga$_2$O$_3$ and 3C-SiC. Similar results have been reported in Si/β-Ga$_2$O$_3$,[20] GaAs/diamond,[26] and InGaP/diamond[27] heterointerfaces.

The TBC value of the 1000 °C-annealed 3C-SiC/β-Ga$_2$O$_3$ interface was found to be a large improvement compared with those of 4H-SiC/β-Ga$_2$O$_3$ interfaces without and with Al$_2$O$_3$ interfacial layer. This should be attributed to the high-quality of the 1000 °C-annealed 3C-SiC/β-Ga$_2$O$_3$ interface. Compared with the amorphous interfacial layers in previous bonded interfaces in the literature, the thin and crystallized interfacial layer (1.5 nm-thick) formed at the 1000 °C-annealed interface facilitates phonon energy transport across the interfaces. Additionally, recent experimental study showed that the intrinsic thermal conductivity of bulk 3C-SiC is over 500 W/m·K, about 50% higher than those of 4H-SiC and 6H-SiC. The corresponding high-quality 3C-SiC thin films have record-high thermal conductivity, even higher than that of diamond thin films with equivalent thicknesses.[28] This motivates us to use 3C-SiC thin films to cool β-Ga$_2$O$_3$ devices from the top-side. The measured thermal conductivity of the 3C-SiC thin film in this work is measured to be 215 W/m·K, which is one order of magnitude higher than that of bulk β-Ga$_2$O$_3$ and close to bulk GaN. The combination of high-quality integrated interfaces with high TBC and high thermal conductivity of 3C-SiC facilitates heat extraction and spreading of related β-Ga$_2$O$_3$ power devices significantly.[29] These results showed that the 3C-SiC/β-Ga$_2$O$_3$ bonding interface has good practicability that withstands all device processes such as β-Ga$_2$O$_3$ crystal growth and ohmic contacts annealing. The strategy demonstrated in this work provides a way for heat extraction and spreading from the top side of both lateral and vertical β-Ga$_2$O$_3$ power devices. The direct bonding of 3C-SiC thin film to the *β*-Ga$_2$O$_3$ substrate can enable better thermal management, which contributes to solve the grand thermal challenge of β-Ga$_2$O$_3$ power electronics.

■ **CONCLUSIONS**

The high thermal conductivity 3C-SiC thin film grown on Si substrate with hole patterns was successfully transferred to *β*-Ga$_2$O$_3$ (001) substrate by the SAB technique at room temperature. A 4.5 nm-thick crystal defect layer was formed at the as-bonded 3C-SiC/*β*-Ga$_2$O$_3$ interface. The crystal defect layer thickness decreased with increasing annealing temperature. The reduction of the crystal layer thickness was attributed to that the crystal defect layer self-healing by the post-annealing process. After annealing at 1000 °C, the crystal defect layer was decreased to 1.5 nm, the TBC value of the 3C-SiC/$\beta$-Ga$_2$O$_3$ interface was measured to be 244 MW/m$^2$·K, which is higher than the previously reported values. The 3C-SiC/$\beta$-Ga$_2$O$_3$ bonding interface has a high thermal stability that can satisfy the requirement of the $\beta$-Ga$_2$O$_3$ device fabrication processes. These results indicated that the high thermal conductivity of 3C-SiC film (215 W/m·K) bonded to $\beta$-Ga$_2$O$_3$ has a large potential for improving the heat extraction and spreading of the $\beta$-Ga$_2$O$_3$-based power devices.

### ■ EXPERIMENTAL SECTION

3C-SiC epitaxial layers grown on Si and $\beta$-Ga$_2$O$_3$ substrates were used for our experiments. The 3C-SiC epitaxial substrate was a ~1-μm-thick 3C-SiC (111) film grown on a Si (111) substrate, which was formed by low-temperature chemical vapor deposition.[28] The averaged roughness (Ra) values of the 3C-SiC layers grown on the Si and the $\beta$-Ga$_2$O$_3$ substrate surfaces were measured to be 0.41 and 0.23 nm, respectively, by an atomic force microscope (AFM) system. Hole patterns were formed on the 3C-SiC epitaxial substrate by photolithography and inductive coupled plasma reactive ion etching processes. And then the 3C-SiC epitaxial layer with hole patterns was directly bonded to the $\beta$-Ga$_2$O$_3$ substrate by the SAB method at room temperature. Detailed bonding processes can be found in our previous works.[30-32] After bonding, the Si substrate was removed by mechanical polishing and reactive ion etching (RIE) process to remain 3C-SiC epitaxial layer on the $\beta$-Ga$_2$O$_3$ substrate.

**Interface analyzation:** The atomic structure and composition distribution of the 3C-SiC/$\beta$-Ga$_2$O$_3$ interface were investigated by TEM (JEM-2200FS) and EDX under STEM with a JEOL JEM-2200F analytical microscope. An acceleration voltage of 200 kV was used. TEM specimens were prepared by the focused ion beam (FIB) technique (Helios NanoLab600i; Thermo Fisher Scientific) at room temperature.

**Thermal Measurements**: The thermal properties of the bonded 3C-SiC-β-Ga$_2$O$_3$ sample are measured by time-domain thermoreflectance (TDTR). TDTR is an ultrafast laser-based pump-probe technique that is able to measure thermal properties of both bulk and nanostructured materials.[33] Before measurements, a layer of ~80-nm-thick Al is deposited on the sample surface as a transducer. A modulated pump laser heats the sample surface periodically while a delayed probe laser detects the temperature variations of the sample surface via thermoreflectance.[34] Here, a 10 X objective is used with a spot size of 5.5 μm (radius) and the modulation frequency is 9.3 MHz.

ACKNOWLEDGMENT

This work was supported by JSPS KAKENHI Grant Number JP20K04581 and the Osaka City University (OCU) Strategic Research Grant 2020 for top basic research. The fabrication of the TEM samples was performed at The Oarai Center and at the Laboratory of Alpha-Ray Emitters in IMR under the Inter-University Cooperative Research in IMR of Tohoku University (NO. 18M0045 and 19M0037). A part of this work was supported by Kyoto University Nano Technology Hub in the "Nanotechnology Platform Project" sponsored by the Ministry of Education, Culture, Sports, Science and Technology (MEXT), Japan. Z. Cheng would like to achnowlede the financial support from U.S. Office of Naval Research under a MURI program (Grant N00014-18-1-2429).